\documentclass[]{aa}

\usepackage[varg]{txfonts}
\usepackage{color}
\usepackage{graphicx}
\usepackage{lscape}
\usepackage{longtable}
\usepackage{natbib}
\usepackage{ulem}
\usepackage{graphicx}
\usepackage{epstopdf}
\usepackage[switch]{lineno}

\bibpunct{(}{)}{;}{a}{}{,} 

\begin{document}

\title{A statistical study of decaying kink oscillations detected using SDO/AIA}
\author{C. R. Goddard \inst{1} \and G. Nistic\`o \inst{1} \and V. M. Nakariakov \inst{1,2,3} \and I. V. Zimovets \inst{4}}
\institute{Centre for Fusion, Space and Astrophysics, Department of Physics, University of Warwick, CV4 7AL, UK, \email{c.r.goddard@warwick.ac.uk}\
\and Astronomical Observatory at Pulkovo of the Russian Academy of Sciences, 196140 St Petersburg, Russia 
\and School of Space Research, Kyung Hee University, 446-701 Yongin, Gyeonggi, Korea
\and Space Research Institute (IKI) of Russian Academy of Sciences, Profsoyuznaya St. 84/32, 117997 Moscow, Russia 
}

\date{Received 10/09/2015, Accepted 04/11/2015}

\abstract 
{Despite intensive studies of kink oscillations of coronal loops in the last decade, a large scale statistically significant investigation of the oscillation parameters has not been made using data from the Solar Dynamics Observatory (SDO).}
{We carry out a statistical study of kink oscillations using Extreme Ultra-Violet (EUV) imaging data from a previously compiled catalogue.}
{We analysed 58 kink oscillation events observed by the Atmospheric Imaging Assembly (AIA) onboard SDO during its first four years of operation (2010-2014). Parameters of the oscillations, including the initial apparent amplitude, period, length of the oscillating loop, and damping are studied for 120 individual loop oscillations.}
{Analysis of the initial loop displacement and oscillation amplitude leads to the conclusion that the initial loop displacement prescribes the initial amplitude of oscillation in general. The period is found to scale with the loop length, and a linear fit of the data cloud gives a kink speed of $C_k$=(1330$\pm$50) km s$^{-1}$. The main body of the data corresponds to kink speeds in the range $C_k$=(800-3300) km s$^{-1}$. Measurements of 52 exponential damping times were made, and it was noted that at least 22 of the damping profiles may be better approximated by a combination of non-exponential and exponential profiles, rather than a purely exponential damping envelope. There are an additional 10 cases where the profile appears to be purely non-exponential, and no damping time was measured.  A scaling of the exponential damping time with the period is found, following the previously established linear scaling between these two parameters. }
{}

\keywords{Sun: corona - Sun: oscillations - methods: observational}

\maketitle

\titlerunning{Kink oscillations - a statistical study}
\authorrunning{Goddard et al.}

\section{Introduction}

	Waves and oscillations in the solar corona have been observed and studied for decades. Detections at Extreme Ultra-Violet (EUV) and X-ray wavelengths have confirmed theories and models related to magnetohydrodynamic (MHD) wave modes in magnetically and gravitationally structured plasmas \citep{2012RSPTA.370.3193D}. The importance of these waves and oscillations stems from their relation to the local plasma parameters of the medium, allowing coronal seismology to be attempted \citep{2012book, 2014SoPh..289.3233L}.
	
	Kink oscillations of coronal loops have been intensively studied since their detection with the Transition Region And Coronal Explorer (TRACE) \citep{1999SoPh..187..229H} in 1999 \citep{1999ApJ...520..880A, 1999Sci...285..862N}. Prior to their detection they were the subject of a range of theoretical and numerical studies \citep[e.g.][]{1982SvAL....8..132Z, 1983SoPh...88..179E, 1984ApJ...279..857R, 1994SoPh..151..305M}.  This phenomenon is clearly observed with the spatial and temporal resolution of recent EUV imagers, such as TRACE, and the Atmospheric Imaging Assembly (AIA) onboard the Solar Dynamics Observatory (SDO) \citep{2012SoPh..275...17L}. Standing global modes induced by flaring activity are the most commonly detected form of kink oscillation \citep[e.g.][]{1999Sci...285..862N}. Other detections have included higher spatial harmonics in coronal loops \citep[e.g.][]{2007A&A...473..959V}, their propagating form \citep{2007Sci...317.1192T}, oscillations of polar plumes \citep{2014ApJ...790L...2T}, propagating kink waves in streamer stalks \citep{2010ApJ...714..644C, 2011ApJ...728..147C} and kink waves in coronal jets \citep{2009A&A...498L..29V}.
	
	The period of a fundamental kink oscillation is given by the expression $P=2L/C_k$, where $L$ is the loop length and $C_k$ is the kink speed, which is the phase speed of the kink wave \citep{1982SvAL....8..132Z, 1983SoPh...88..179E}. The kink speed is dependant on the internal Alfv\'en speed ($C_{A0}$) and the density contrast between the ambient plasma and the loop ($\rho_e/\rho_0$), in the low-$\beta$ plasma limit this can be approximated as $C_k=(2/(1+\rho_e/\rho_0))^{1/2} C_{A0}$. The Alfv\'en speed is given by $C_{A0}=B_0/\sqrt{\mu_0 \rho_0}$, meaning that if a kink speed is measured and the density contrast can be estimated then the magnetic field in the loop, $B_0$, can be estimated \citep[e.g.][]{2001A&A...372L..53N}. 
	
	Standing kink oscillations can be induced by a flare-generated blast wave \citep[e.g.][]{2008ApJ...682.1338M}, by periodic driving from a coronal wave \citep{2010A&A...512A..76S}, or loop contraction due to the reconfiguration of the active region magnetic field after an eruption \citep{2015A&A...581A...8R}. Alternative mechanisms such as Alfv\'enic vortex shedding \citep{2009A&A...502..661N} have also been proposed. However, it was recently shown that the vast majority of kink oscillations of coronal loops are excited by low coronal eruptions (LCE) \citep{2015A&A...577A...4Z}, discussed in more detail below.
		
	The commonly applied explanation for the observed rapid damping of these oscillations is resonant absorption. The wave is considered to be a kink mode which evolves, through a resonance at a certain thin loop layer, from a global kink mode to a local azimuthal motion \citep[e.g.][]{2002ApJ...577..475R, 2002A&A...394L..39G}. The local motion cannot be observed directly and it should undergo its own damping. Other proposed mechanisms include phase mixing \citep[e.g.][]{2002ApJ...576L.153O}, and viscous or dissipative damping \citep[e.g.][]{1999Sci...285..862N}.
	
	For standing kink modes, it has been shown that an exponential damping profile is expected \citep{2002ApJ...577..475R}. Numerical simulations from \cite{2012A&A...539A..37P} revealed a change in the characteristic damping profile of propagating transverse velocity displacements, which is directly applicable to standing kink modes. At the beginning of the oscillation a Gaussian profile is present, followed by an exponential damping profile, the time at which the transition between the two damping profiles occurs was suggested to depend on the transverse density structuring. Purely exponential, purely Gaussian, and a Gaussian followed by exponential damping are possible depending on the parameters of the loop and background plasma. More recently \cite{2013A&A...551A..40P} performed an analytical and parametric study of kink wave damping via mode coupling, which confirmed the existence of a Gaussian phase in the damping envelope, which was found to be more suitable for seismological application than the asymptotic exponential stage. 

	 \cite{2002ApJ...576L.153O} studied scaling relations for different damping mechanisms, and determined that the observations considered were consistent with the phase mixing mechanism \citep{1983A&A...117..220H}. However, as discussed by \cite{2008ApJ...676L..77A}, a comparison between the observed scaling and the linear scaling typical for resonant absorption requires that all loops have the same cross-sectional structuring, and accounting for this the scaling from resonant absorption can depart from a linear dependence between damping time and period.
	
	There have been various applications of kink oscillation observations to perform remote diagnostics of the parameters of the surrounding plasma \citep[e.g.][]{2012PhyU...55A...4S}. Through comparisons to modelling results the absolute value of the loops internal magnetic field has been estimated \citep[e.g.][]{2001A&A...372L..53N}, the density scale height within the loop can be inferred \citep[e.g.][]{2005ApJ...624L..57A} as well as characterisation of the height variation of the loop minor radius \citep[e.g.][]{2008A&A...486.1015V}.
		
	Recent observational studies of kink oscillations in the corona include \cite{2013A&A...552A.138V}, where results were combined to study the scaling of the damping time with period, and put constraints on the loop’s density contrast and inhomogeneity layer thickness. In addition, a decayless regime of kink oscillations has been detected \citep{2013A&A...552A..57N, 2013A&A...560A.107A} and the statistical properties of this regime have been determined \citep{2015temp}. A linear scaling of the period with loop length was obtained, confirming their interpretation of the oscillations as standing kink waves. Recently \cite{2015A&A...577A...4Z} catalogued 169 individual decaying kink oscillations using SDO/AIA data, and utilised these observations to shed light on the excitation mechanism for the oscillations. They found that 95$\%$ of the oscillations corresponded to displacement of the loop by a nearby lower coronal eruption/ejection, indicating this is the most common excitation mechanism. An LCE is the emergence of an expanding plasma structure (such as a flux rope, filament, jet or loop arcade) or plasma ejection (mainly along magnetic field lines) into the lower corona (roughly $R<1.4R_{\odot}$), usually, but not necessarily, accompanied by a flare and a subsequent CME.
	
	The aim of our study is to use this catalogue to investigate the distributions of the parameters describing the initial excitation of each oscillation, the oscillation itself and its damping. The determination of statistical relationships between the different parameters of kink oscillations of coronal loops is also performed. In Section 2 the data and analysis are described, in Section 3 the results are presented, and a discussion and conclusion are given in Section 4.


\begin{figure}
\resizebox{\hsize}{!}{\includegraphics{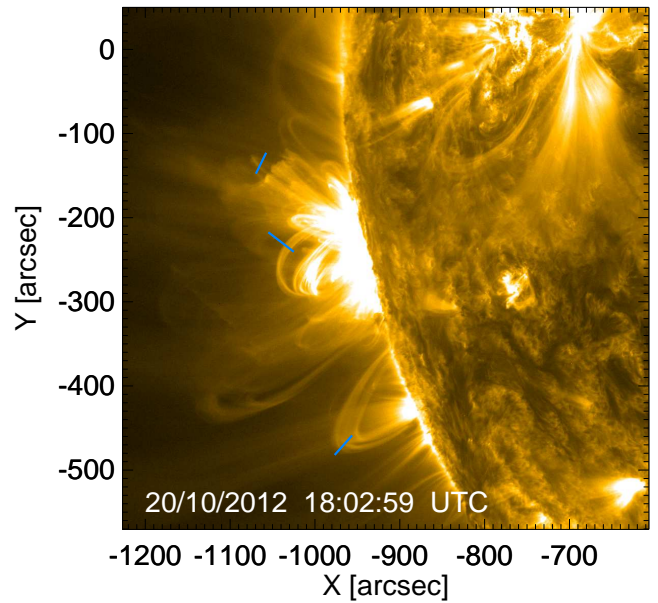}}
\caption{ The active region from event number 40 from Table \ref{table1}. The three blue lines show some of the slits used to create time-distance maps for analysing the oscillations of the corresponding loops. }
\label{ar_fig}
\end{figure}

\begin{figure}
\resizebox{\hsize}{!}{\includegraphics{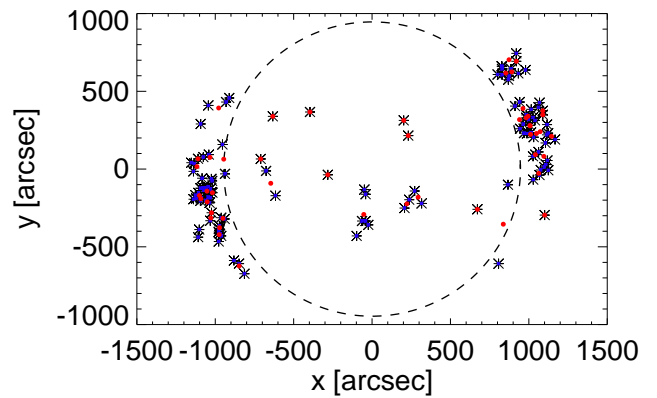}}
\caption{ The slit positions (x1, y1 from Table \ref{table1}) used to produce the sample of time-distance maps to analyse kink oscillations of coronal loops, plotted as blue asterisks. The overplotted red circles are the average slit position for each event. }
\label{pos_fig}
\end{figure}

\begin{figure}
\resizebox{\hsize}{!}{\includegraphics{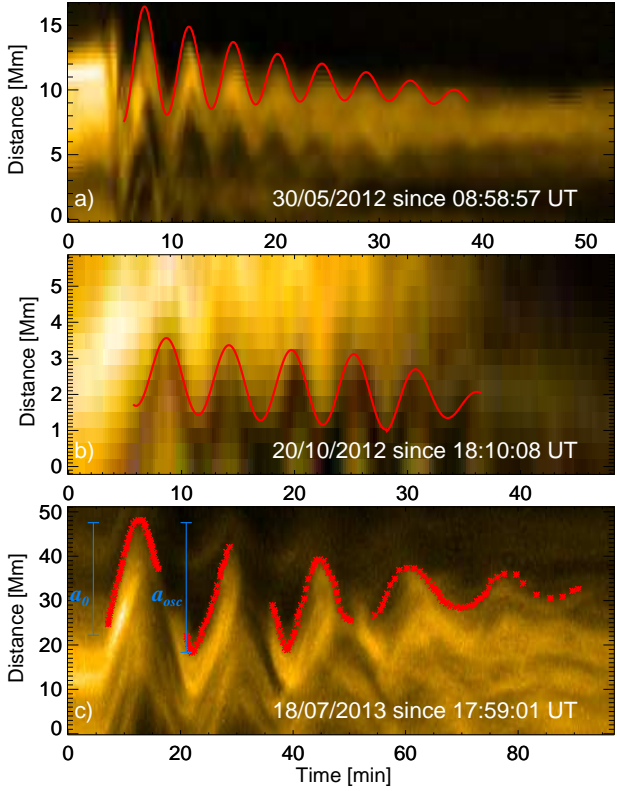}}
\caption{ Three typical time distance maps, corresponding to loop 1 from event 32, loop 4 from event 40, and loop 2 from event 48. The fits correspond to the detrending polynomial added to the sinusoidal fit, and multiplied by the exponential damping profile for panels a) and b). The red points in panel c) were taken by eye to map out the oscillation and used for the subsequent fitting. The vertical blue bars represent the measurement of the initial displacement ($a_0$) and the initial oscillation amplitude ($a_{osc}$) of the upper loop edge.}
\label{td_maps}
\end{figure}

\section{Observations and analysis}

	Decaying kink oscillations of coronal loops from the catalogue in \cite{2015A&A...577A...4Z} are analysed using 171 $\AA$ data from SDO/AIA. A series of images was obtained for each of the 58 kink oscillation events listed using the provided date, oscillation time and location to define a field of view and time interval, which were subsequently submitted as a SolarSoft (SSW) cut-out request. The data cubes obtained had a time span of 30 or 45 minutes (which was extended for long period oscillations as required), the standard pixel size of 0.6 arcsec, and a temporal cadence of 12 s, or 24 s in some cases, when the intermediate image was not returned due to a short exposure time.
	
	Movies created from the data cubes were initially inspected by eye, and loop oscillations with sufficient quality for time-distance (TD) analysis were noted. TD maps were created by taking linear slits with a 5 pixel width perpendicular to the oscillating loop and stacking the intensities along the slit (averaged over the width to increase the signal-to-noise ratio) in time. An example of an active region is show in Fig. \ref{ar_fig}, with the slits used to create TD maps of different loops overplotted. This process resulted in 127 TD maps, which were interpolated to an equispaced temporal grid of 12 s. The ends of the slits used are listed in Table \ref{table1}, along with the event number from \cite{2015A&A...577A...4Z} and a loop ID. The first end-point of each slit is overplotted on the disk in Fig. \ref{pos_fig} in blue, and the average slit position for each event is plotted in red, the observed clustering is due to different events occurring within the same active region.
	
	The projected loop length was estimated for each oscillating loop. The major radius (via the apparent loop height) or diameter (via the distance between footpoints) was measured by eye, depending on the orientation of the loop with respect to our line of sight (LoS), and a semicircular loop approximation was used, $L=\pi R$. The loop lengths are listed in Table \ref{table1}. In a few cases the loop length could not be estimated as the footpoint positions or height could not be determined.
	
	For each TD map the amplitude of the initial displacement and initial oscillation amplitude were recorded. The initial displacement is defined as the difference between the initial loop position and the first maximum, and the initial amplitude is defined between the first maximum and minimum, as shown in Fig. \ref{td_maps} c). In this case the displacement of the loops upper edge was estimated. The start time of the oscillation was also recorded, in addition to the number of oscillation cycles observed, all listed in Table \ref{table1}.
	
	Gaussian tracking of the loop centre to record the oscillating displacement was not appropriate in many cases due to the overlap of multiple loops, or only the edge of the oscillating loop being clearly defined. Due to this the oscillations were mapped out by taking a series of points along the centre or edge of the loops by hand, and an error for the points was defined based on the clarity of the time distance map, $\pm$ 1 pixel in most cases. Three examples of TD maps are shown in Fig. \ref{td_maps}, the points used to define the oscillation are overplotted in panel Fig. \ref{td_maps} c), where the loop edge was mapped out and points were not taken when its position could not be reliably determined.
	
	There is a certain degree of subjectiveness and error associated with the loop length measurement, the displacement measurements and the points taken to map out the oscillation, however the sample size is large enough that this will not affect the overall results, and only affects individual measurements.
	
	The data points for each oscillation were detrended by fitting with a polynomial function of the form  $ y= C_0 + C_1t + C_2t^{2} + C_3t^{3} $, and subtracting this from the data. All fitting was performed with user defined functions and the IDL rountine \texttt{mpfitexpr.pro}. Fitting with a sinusoidal function, of the form $y = A \sin({2\pi}t/P + \phi)$, was performed for each detrended oscillation, with the period ($P$) as one of the free parameters. The best fitting period and the corresponding error were recorded for each well defined oscillation.
	
	To analyse the damping behaviour of the oscillations the absolute value of the detrended oscillatory signal was taken and scrutinised by eye. For \textgreater 50$\%$ of the TD maps clear damping could not be seen, or the number of oscillation cycles was not sufficient to perform fitting of the damping envelope. However these oscillations are still clearly part of the decaying, rather than decayless, regime. For oscillations with a clear exponential decaying trend a weighted fit, of the form $A(t)=A_0e^{-t/\tau}$, was performed on the maxima of the absolute value of the detrended signal. The damping time $\tau$ and the corresponding error were recorded. For cases where the damping was not observed for the whole duration of the signal, or where there were clearly non-exponential regions of the damping envelope, the fit was only made for the region which was approximated well by an exponential decay. By eye it was determined whether each damping profile was best described by a purely exponential profile (see Fig. \ref{td_maps} panel a) ), or a combination of both non-exponential and  exponential profiles (see Fig. \ref{td_maps} panel b) ).
	
	The result of the whole fitting process is overplotted on the TD maps in Fig. \ref{td_maps}, where the detrending polynomial has been added to the sinusoidal fit, as well as the damping profile for panels a) and b). For panel a) the damping profile was measured for the whole signal, for panel b) it was only measured for the last 2 cycles.

\section{Results}

\begin{figure}
\resizebox{\hsize}{!}{\includegraphics{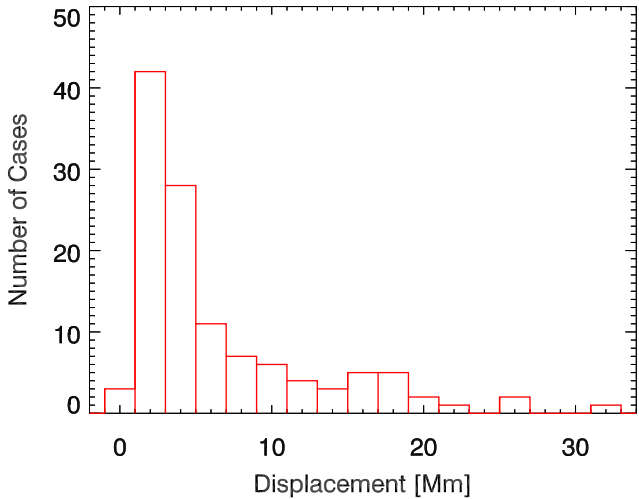}}
\caption{ The distribution of the measured initial displacement of 120 kink oscillations of coronal loops. The bin size is 2 Mm.}
\label{disp_hist}
\end{figure}

\begin{figure}
\resizebox{\hsize}{!}{\includegraphics{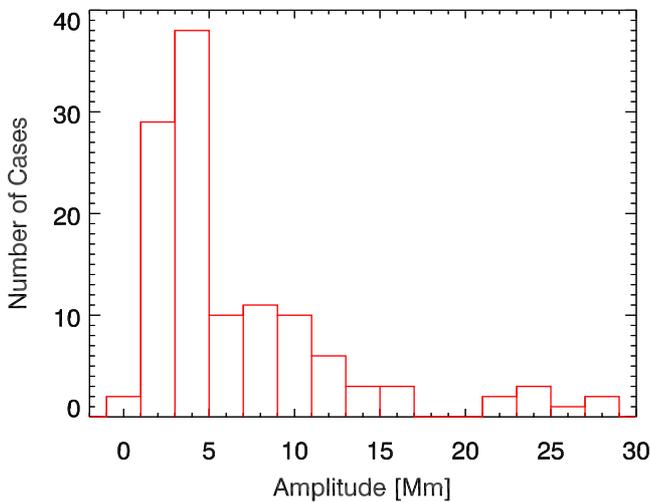}}
\caption{ The distribution of the measured initial oscillation amplitude of 120 kink oscillations of coronal loops, recorded from the first cycle of oscillation after the initial displacement. The bin size is 2 Mm.}
\label{amp_hist}
\end{figure}

\begin{figure}
\resizebox{\hsize}{!}{\includegraphics{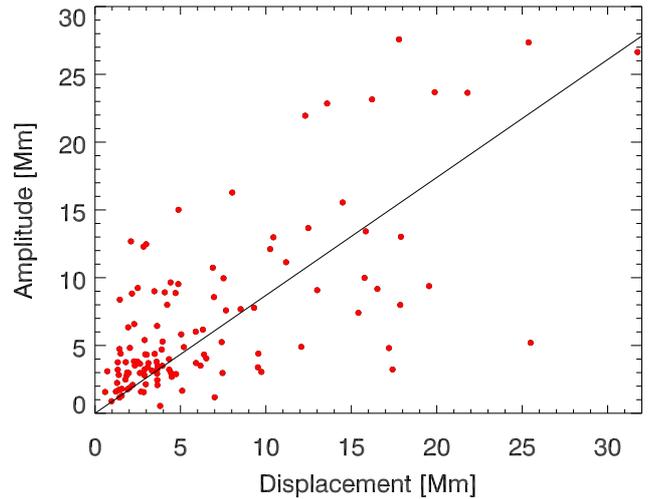}}
\caption{ The initial oscillation amplitude of 120 kink oscillations of coronal loops, plotted against the initial displacement of the loop position, both measured in Mm. A linear fit of the data passing through the origin is shown by the solid black line, with a gradient of 0.86 $\pm$ 0.01. }
\label{amp_plot}
\end{figure}

	From the 58 events analysed the periods of 120 individual kink oscillations were determined, and for 118 of these the corresponding loop length was estimated. From the oscillatory signals 52 exponential damping times were obtained. In addition to these measurements other details of the oscillations were recorded, and are listed in Table \ref{table1}. This included whether damping was observed, and if so whether it is best described by a purely exponential, or a combination of non-exponential and exponential damping profiles. The number of oscillation cycles was also recorded. In many cases this number was limited by the oscillatory signal becoming unclear, and not the damping reducing the oscillation to an undetectable amplitude. 

\subsection{Oscillation parameter histograms}

	Analysis of the amplitudes of the initial displacement and subsequent oscillation may allow inferences to be made about the excitation mechanism, as well as giving details of the typical scales involved. The measured initial displacements ranged from 0.6 to 31.8 Mm. In Fig. \ref{disp_hist} a histogram of the measured initial loop displacements is shown, with a bin size of 2 Mm. The distribution peaks strongly at 1--3 Mm, and 59$\%$ of the measurements are covered by the range 1--5 Mm. The distribution after this range is more uniform, but the number of cases decreases towards the upper limit. 
	
	The initial oscillation amplitudes ranged from 0.5 to 27.6 Mm. In Fig. \ref{amp_hist} a histogram of the measured initial oscillation amplitudes is shown, with a bin size of 2 Mm. The distribution peaks at 3--5 Mm, with 57$\%$ of the measurements lying in the range 1--5 Mm. The distribution again flattens and decreases towards the upper limit of the measurements.
		
	In Fig. \ref{amp_plot} the initial oscillation amplitude is plotted against the initial loop displacement. A rough correlation between these two parameters is observed.  A linear fit of the data cloud which passes through the origin gives a gradient of 0.86 $\pm$ 0.01.
	
		The measured oscillation periods ranged from 1.5 to 28 min. In Fig. \ref{period_hist} a histogram of the measured oscillation periods is shown.  The distribution peaks at 4--7 min, and drops quickly to the maximum detected period, 28 min (not shown in the histogram). No periods below 1.5 min were recorded, and there is a decrease in occurrence approaching the lower periods. In Fig. \ref{length_hist} a histogram of the measured loop lengths is shown. The most common length is in the range 220--260 Mm, but there is a roughly even distribution between 140 and 460 Mm which decreases above and below this range, with minimum and maximum values of 77 and 596 Mm, respectively. 
	
\subsection{Dependence of the period on length}
	
	In Fig. \ref{period_plot} the period is plotted against the loop length, and the period clearly increases with the length.  The period errors correspond to the scaled covariance from the period fitting. An unweighed linear fit was made, as the errors alone do not reflect the distribution of the data. The variation of the density contrast and Alfv\'en speed between different loops and active regions also introduce an intrinsic spread to the data.
		
	The black line corresponds to an unweighted linear fit of the data. The best fitting linear function is $P$[min]=(0.025$\pm$0.001)$L$[Mm], where $L$ is the loop length and $P$ is the period, giving a kink speed of $C_k$=(1330$\pm$50) km s$^{-1}$ from the empirically determined gradient and the equation $P=2L/C_k$. The gradient can be varied to give upper and lower bounds to the data cloud, giving a kink speed range of $C_k$=(800--3300) km s$^{-1}$. Calculation of the kink speed for each individual data point gives the distribution shown in the inset histogram. This has a most common value of 900--1100 km s$^{-1}$, a peak value of of 1340 $\pm$ 60 km s$^{-1}$ and a Gaussian width of 620 $\pm$ 60 km s$^{-1}$, the later two values were obtained by fitting the observed distribution with a Gaussian model. 

\subsection{Relationship between the damping time and period}
		
		In Fig. \ref{tau_plot} the damping time is plotted against the period, and a correlation between the two parameters is obtained.  The damping time errors correspond to the scaled covariance from the exponential damping fits. The statistics are limited to the cases where the damping time could be measured (see Table \ref{table1}), so the figure is less populated than Fig. \ref{period_plot}. A weighed linear fit was made, corresponding to the solid black line. The best fitting linear function is $\tau$=(1.53$\pm$0.03)$P$, where $\tau$ is the damping time and $P$ is the period.
		
		In Fig. \ref{tau_plot} the red circles correspond to damping profiles which were determined to be exponential by eye, and the blue squares correspond to damping profiles best described by a combination of a non-exponential and exponential profile. For the latter case, which corresponds to 22 of the measurements, the damping time is determined from the exponential part of the profile. This corresponds to the values \lq\lq E \rq\rq and \lq\lq E,NE \rq\rq in the column \lq\lq Damping Profile \rq\rq in Table \ref{table1}. No difference between the two cases is observed. There are 10 additional cases where the profile appears to be purely non-exponential, and no damping time was measured, noted in Table \ref{table1} by \lq\lq NE \rq\rq.

\begin{figure}
\resizebox{\hsize}{!}{\includegraphics{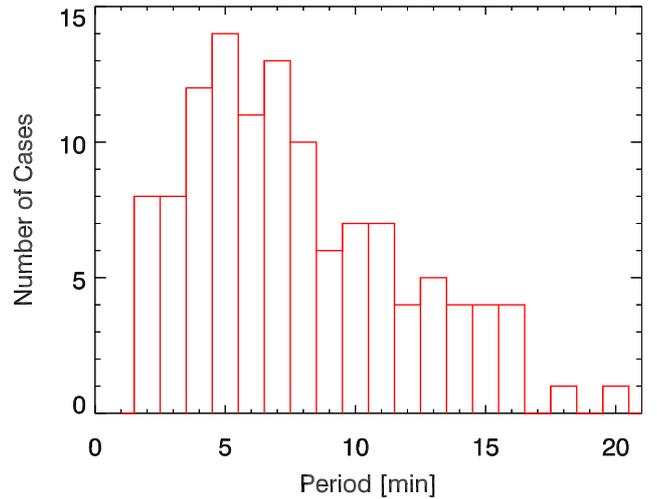}}
\caption{ The distribution of the measured periods of 120 individual kink oscillations of coronal loops. The bin size is 1 min.}
\label{period_hist}
\end{figure}

\begin{figure}
\resizebox{\hsize}{!}{\includegraphics{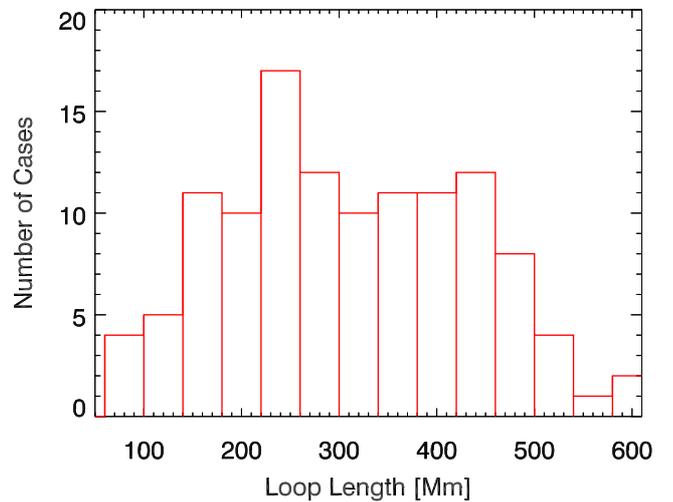}}
\caption{ The distribution of the measured loop lengths for 118 individual coronal loops, which undergo kink oscillations. The bin size is 20 Mm.}
\label{length_hist}
\end{figure}

\begin{figure}
\resizebox{\hsize}{!}{\includegraphics{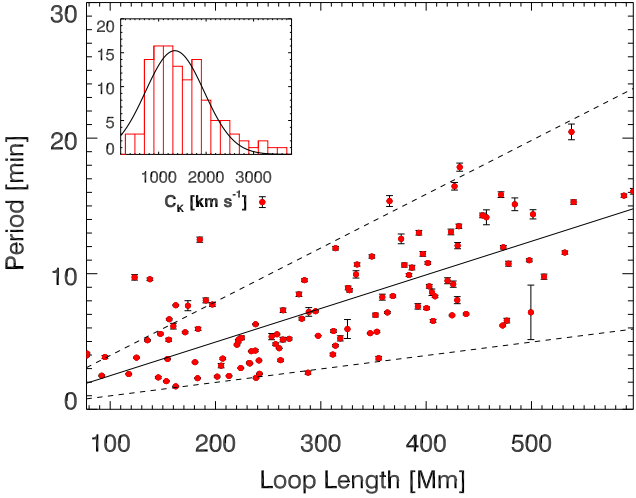}}
\caption{ Period plotted against loop length for 118 kink oscillations of coronal loops. The solid black line correspond to an unweighted linear fit of the data. The best fitting linear function is $P$[min]=(0.025$\pm$0.001)$L$[Mm], giving a kink speed of $C_k$=(1330$\pm$50)km s$^{-1}$ from the gradient. The dashed lines correspond to kink speeds of 800 and 3300 km s$^{-1}$ for the upper and lower lines respectively.}
\label{period_plot}
\end{figure}

\begin{figure}
\resizebox{\hsize}{!}{\includegraphics{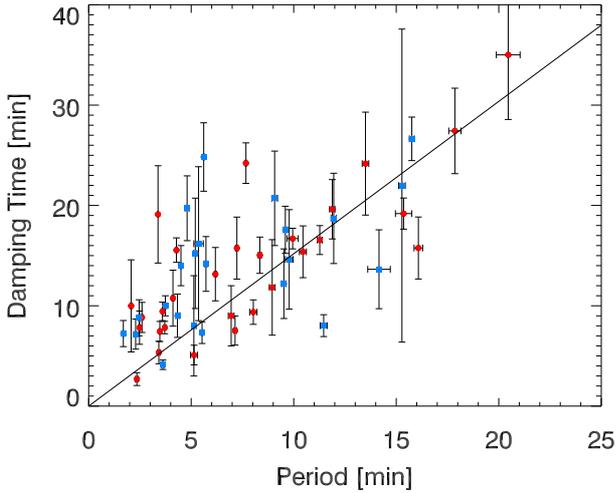}}
\caption{ Damping time plotted against period for 54 kink oscillations of coronal loops. The solid black line correspond to a weighted linear fit of the data. The best fitting linear function is $\tau$=(1.53$\pm$0.03)$P$. The red circles correspond to damping envelopes best described by an exponential profile, and the blue squares correspond to those best described by a combination of a non-exponential and exponential.}
\label{tau_plot}
\end{figure}


\section{Discussion and conclusions}

	In this paper we provide a comprehensive statistical analysis of decaying kink oscillations of coronal loops excited by flaring events, observed with SDO/AIA at 171 $\AA$.

\subsection{Oscillation parameter histograms}

	Details of the initial loop displacement and the subsequent oscillation have been analysed. Fig. \ref{disp_hist} indicates that initiation of kink oscillations by lower amplitude initial displacements is more common, and Fig. \ref{amp_hist} shows that relatively low initial oscillation amplitudes are also more common. A comparison between the two histograms mentioned leads to the conclusion that the initial displacement prescribes the initial amplitude of oscillation in the majority of cases. However, there are cases where a large initial displacement leads to a new equilibrium position for the loop, where it oscillates with an amplitude much smaller than the initial displacement, as recently discussed by \cite{2015A&A...581A...8R}, but the exact cause of this effect still remains unknown. Cases where a small initial displacement results in a larger amplitude oscillation are also seen. A simple explanation is that the passage of the LCE has left the external plasma and magnetic field more rarefied, and the perturbed loop passes through its initial equilibrium with an amplitude grater than the initial displacement. A linear correlation between these two parameters is seen in Fig. \ref{amp_plot}, with spreading due to measurement errors, but also reflecting the different regimes discussed above. 
	
	 The above results are subject to a LoS effect. The histograms in Fig. 3 and 4 will include an effect in the distribution relating to how the observed kink oscillations are distributed over the LoS angles. If all the initial displacements and oscillation amplitudes were equal, distributions would still be obtained due to the different LoS angles varying the measured values. It can be seen from Fig. \ref{pos_fig} that loop positions are mostly off-limb, near the equator, so North-South oscillation polarisations were better detected. Further work is required to remove the effect of the varying LoS angles from the measurements and obtain the true displacements and oscillation amplitudes.
	 
	Some inferences can be made from the period distribution shown in Fig. \ref{period_hist}. The drop off in the occurrence of higher periods is likely to reflect a physical drop-off, as larger length loops may be less likely to be formed and are also more difficult to detect. The  decrease in the distribution for lower periods may include an observational bias, as oscillations of shorter loops are more difficult to observe, but may also reflect the excitation mechanism. If, as proposed by \cite{2015A&A...577A...4Z}, excitation due to LCE perturbations is the dominant mechanism, this should preferentially excite loops above a certain height, which have correspondingly longer lengths. The distributions of the periods and loop lengths differ, and this is likely to be due to the loop length estimations becoming more inaccurate for longer loops, as well as variation of the additional parameters which prescribe the period of oscillation.
	
	If general statistics of the loop lengths were available it could be used to normalise the loop length distribution in Fig. \ref{length_hist}.  This would allow us to determine whether kink oscillations occur in all loops with the same probability or if loops of certain lengths are more likely to undergo a kink oscillation. LoS effects should also be included in such a study, and the varying ellipticity of the loops themselves. 
		
		\subsection{Dependence of the period on length}
			
	The period scales with the loop length as expected, and the best fitting kink speed of $C_k$=(1330$\pm$50) km s$^{-1}$ is in agreement with previous results. The range of kink speeds from the main data cloud ($C_k$=(800--3300) km s$^{-1}$) should correspond to the spread of loop density contrasts, transverse density profiles and Alfv\'en speeds. There are some points in Fig. \ref{period_plot} corresponding to short loops with larger periods than expected from the main body of the data, giving a lower kink speed value. These data points may correspond to loops and active regions with significantly lower Alfv\'en speeds, or greater density contrasts. More statistics and analysis of the active region parameters are required to determine if these points have a physical explanation, or are due to measurement errors and/or random spread.
	
		Comparisons can be made with the results of a similar study which focussed on the decayless regime of kink oscillations \citep{2015temp}. They obtained a similar scaling of period with loop length, but with a lower gradient, and therefore, a higher best fitting kink speed of $C_k$=(1850$\pm$70) km s$^{-1}$. The period distribution they present is similar to the distribution we obtain, but peaks at a lower value, and their loop length distribution is significantly less uniform than ours. These differences may be due to a selection effect from their study, as the data spanned 1 month, whereas the data presented here span 4 years. It may also reflect the different driving mechanisms, in particular the excitation mechanism for the decayless regime remains unknown. The discrepancy between the statistics of decaying and decayless kink oscillations should be further investigated when a larger set of events becomes available. 
	
	\subsection{Relationship between the damping time and period}
	
	Our results on the linear scaling of the damping time with the oscillation period are mainly consistent with the previously obtained results \citep{2002SoPh..206...99A, 2013A&A...552A.138V}. The spread of the data makes it unreasonable to make inferences about the damping mechanism from the observed dependence. In particular it is not possible to discriminate between a linear or power law dependence, and such an approach requires consideration of the influence of the varying cross sectional loop structuring and other parameters. 
	
		Distinguishing between the exponential damping time measured from oscillations which showed a purely exponential profile and those with a combination of a non-exponential and exponential profiles did not reveal any systematic difference. Only the exponential region was fit if there was also a non-exponential region present. This indicates that the exponential stage of the damping is the same whether a non-exponential stage is present or not. Our findings indicate that in some cases the exponential fitting of the kink oscillation damping is not sufficient to capture the whole damping profile. The amplitude of some oscillations remains approximately constant for a significant period of time, it has not been determined whether this corresponds to a slowly decaying Gaussian profile, or a periodic driver sustaining a constant amplitude oscillation.
	
	The detected clearly non-exponential sections of the damping envelopes may be better approximated by a Gaussian profile. If this is confirmed to be the case then this is evidence for the Gaussian damping regime discussed by \cite{2012A&A...539A..37P}. This will be the subject of further study, as the detection of Gaussian damping envelopes allows new seismology to be performed, and comparisons with the theoretical predictions to be made.

\subsection{Conclusion}

We summarise our main findings as follows;
\begin{itemize}	
	\item The initial loop displacement prescribes the initial oscillation amplitude in the majority of cases.
	\item The period scales linearly with the loop length, as expected, and a kink speed of $C_k$=(1330$\pm$50) km s$^{-1}$ is obtained, with the majority of the data points lying in the range (800--3300) km s$^{-1}$, following a Gaussian distribution.
	\item A linear scaling of the damping time with period is observed, and non-exponential damping profiles have been detected.
\end{itemize}

	In conclusion, a statistically significant number of individual kink oscillations have been analysed, and histograms of the measured parameters have allowed insightful inferences to be made. Details of the distribution of amplitudes, periods and loop lengths may be useful when considering the observational capabilities of future instruments. In addition the scaling between different parameters has been studied, and the damping behaviour has been characterised, both of which, after further work, may allow seismological inferences and measurements to be made.

\begin{acknowledgements}
The work was supported by the European Research Council under the SeismoSun Research Project No. 321141 (CRG, VMN), and the STFC consolidated grant ST/L000733/1 (GN, VMN). IVZ was partially supported by the RFBR (research project No. 15-32-21078). The data is used courtesy of the SDO/AIA team. 
\end{acknowledgements}

\bibliographystyle{aa} 
\bibliography{references} 

\onecolumn
\begin{landscape}
\begin{longtable}{|l|l|l|l|l|l|l|l|l|l|l|l|l|}
\caption{A list of 120 coronal loop kink oscillations detected with AIA/SDO and their measured parameters. The event ID corresponds to the events catalogued in \cite{2015A&A...577A...4Z}, and the loop ID distinguishes the different loops in each event (this does not correspond to those in the cited paper). The position of the slit used to produce each time-distance map is given in arcsec, along with the date and oscillation start time in UT. The period and error obtained from fitting the loop oscillation are given, as well as the estimated loop length. The column \lq\lq Disp Amp \rq\rq  lists the estimated initial loop displacement, and \lq\lq Osc Amp \rq\rq  is the estimated initial amplitude of the oscillation. The number of cycles that were observed is listed in \lq\lq N Cyc \rq\rq. Finally, the exponential damping time and error from fitting the damping profile, and the form of the damping profile (exponential (E), non-exponential(NE), or a combination of both), are listed in the final two columns.}
\\
\hline
Event & Loop & Slit Position         & Date       & Time     & Period         & Length & Disp Amp & Osc Amp & N Cyc & Damping Time   & Damping  \\
ID    & ID   & [x1,y1,x2,y2] (arcsec)&            & UT       & (min)          & (Mm)   & (Mm)     & (Mm)    &       & (min)          & Profile  \\
\hline
1     & 1    & -940,-321,-964,-308   & 02/08/2010 & 04:22:49 & 3.42$\pm$0.06  & 232    & 5.1      & 1.7     & 3     & 5.34$\pm$1.12  & E        \\
1     & 2    & -962,-313,-997,-322   & 02/08/2010 & 04:22:13 & 4.11$\pm$0.05  & 78     & 7.0      & 1.2     & 3     & 10.76$\pm$2.79 & E        \\
2     & 1    & 672,-259,711,-223     & 16/10/2010 & 19:13:07 & 6.64$\pm$0.06  & 156    & 2.0      & 4.8     & 3.5   &                &          \\
3     & 1    & -977,-383,-988,-368   & 03/11/2010 & 12:13:48 & 2.46$\pm$0.03  & 213    & 1.4      & 4.7     & 8     & 8.8$\pm$1.8    & E,NE     \\
3     & 2    & -970,-416,-1001,-393  & 03/11/2010 & 12:14:35 & 3.62$\pm$0.08  & 262    & 4.4      & 9.7     & 3     & 4.12$\pm$0.47  & E,NE     \\
3     & 3    & -978,-466,-1027,-411  & 03/11/2010 & 12:14:23 & 4.04$\pm$0.1   & 311    & 4.1      & 8.9     & 2     &                &          \\
4     & 1    & 912,405,889,433       & 09/02/2011 & 01:30:02 & 2.29$\pm$0.03  & 183    & 2.9      & 4.4     & 4.5   & 7.18$\pm$1.5   & E,NE     \\
4     & 2    & 969,231,974,278       & 09/02/2011 & 01:31:54 & 3.47$\pm$0.03  & 181    & 1.4      & 1.2     & 3     & 7.44$\pm$1     & E        \\
5     & 1    & 1089,375,1050,423     & 10/02/2011 & 04:43:38 & 7.03$\pm$0.06  & 438    & 4.5      & 3.0     & 3     &                & NE       \\
6     & 1    & 1089,349,1057,398     & 10/02/2011 & 06:44:22 & 8.05$\pm$0.26  & 430    & 3.8      & 0.5     & 2     &                &          \\
7     & 1    & 983,330,970,342       & 10/02/2011 & 06:57:46 & 1.69$\pm$0.02  & 162    & 2.9      & 3.2     & 6     & 7.23$\pm$1.3   & E,NE     \\
8     & 1    & 1007,280,1021,305     & 10/02/2011 & 12:35:01 & 3.74$\pm$0.07  & 207    & 1.2      & 1.6     & 3     & 10$\pm$1       & E,NE     \\
9     & 1    & 983,348,947,414       & 10/02/2011 & 13:43:37 & 5.14$\pm$0.17  & 264    & 3.0      & 4.3     & 3     & 5.09$\pm$0.98  & E        \\
9     & 2    & 942,431,934,461       & 10/02/2011 & 13:46:31 & 8.95$\pm$0.14  & 326    & 3.6      & 3.2     & 2.5   & 11.83$\pm$4.76 & E        \\
10    & 1    & 1106,168,1133,214     & 11/02/2011 & 08:07:07 & 11.46$\pm$0.17 & 397    & 4.7      & 8.9     & 2.5   & 8.02$\pm$1.09  & E,NE     \\
10    & 2    & 1039,313,1041,334     & 11/02/2011 & 08:08:17 & 8.48$\pm$0.16  & 279    & 5.9      & 6.0     & 2     &                &          \\
11    & 1    & -41,-162,-43,-146     & 13/02/2011 & 17:34:28 & 3.96$\pm$0.07  & 78     & 3.5      & 4.4     & 3     &                &          \\
11    & 2    & -49,-132,-51,-108     & 13/02/2011 & 17:34:50 & 3.85$\pm$0.11  & 95     & 3.7      & 2.1     & 3     &                &          \\
11    & 3    & -64,-334,-69,-316     & 13/02/2011 & 17:37:13 & 2.6$\pm$0.05   & 118    & 3.1      & 3.7     & 6     & 8.84$\pm$1.5   & E        \\
11    & 4    & -41,-334,-54,-322     & 13/02/2011 & 17:33:52 & 3.81$\pm$0.04  & 125    & 2.9      & 5.4     & 5     &                &          \\
11    & 5    & -24,-359,-44,-336     & 13/02/2011 & 17:33:42 & 5.09$\pm$0.06  & 135    & 1.9      & 6.3     & 2     &                &          \\
11    & 6    & -98,-430,-89,-394     & 13/02/2011 & 17:38:33 & 6.13$\pm$0.21  & 160    & 11.2     & 11.1    & 2     &                &          \\
12    & 1    & -282,-37,-309,-47     & 13/02/2011 & 20:19:17 & 5.56$\pm$0.07  & 148    & 1.9      & 1.8     & 2     &                &          \\
15    & 1    & 202,313,175,371       & 27/05/2011 & 10:47:58 & 7.64$\pm$0.37  & 174    & 6.3      & 6.2     & 1.5   &                &          \\
16    & 1    & 1014,235,991,257      & 11/08/2011 & 10:17:19 & 2.62$\pm$0.04  & 242    & 3.3      & 3.1     & 3     &                &          \\
16    & 2    & 988,229,1026,229      & 11/08/2011 & 10:10:22 & 2.35$\pm$0.07  & 146    & 17.4     & 3.2     & 2     & 2.69$\pm$0.64  & E        \\
16    & 3    & 1031,205,1067,241     & 11/08/2011 & 10:10:54 & 5.23$\pm$0.19  & 318    & 25.5     & 5.2     & 2.5   &                &          \\
17    & 1    & 231,215,216,263       & 06/09/2001 & 22:20:15 & 2.07$\pm$0.04  & 153    & 9.5      & 3.4     & 3.5   & 9.99$\pm$4.59  & E        \\
18    & 1    & -931,431,-960,472     & 22/09/2011 & 10:35:08 & 7.18$\pm$0.32  & 289    & 15.8     & 10.0    & 2.5   &                &          \\
18    & 2    & -911,457,-884,476     & 22/09/2011 & 10:26:59 & 9.52$\pm$0.11  & 284    & 1.4      & 1.7     & 3.5   & 12.2$\pm$3.47  & E        \\
18    & 3    & -1093,290,-1060,320   & 22/09/2011 & 10:30:32 & 13.02$\pm$0.17 & 393    & 4.9      & 9.5     & 4     &                & NE       \\
19    & 1    & -954,158,-998,134     & 23/09/2011 & 23:51:45 & 9.73$\pm$0.2   & 123    & 4.7      & 2.9     & 2     &                &          \\
19    & 2    & -938,-31,-992,-12     & 23/09/2011 & 23:51:57 & 11.27$\pm$0.12 & 348    & 7.5      & 10.0    & 2     & 16.55$\pm$1.44 & E        \\
20    & 1    & -676,-12,-682,62      & 14/11/2011 & 07:21:12 & 5.36$\pm$0.23  & 253    & 2.6      & 3.7     & 3     & 16.19$\pm$7.67 & E,NE     \\
20    & 2    & -616,-171,-665,-161   & 14/11/2011 & 00:05:04 & 13.43$\pm$0.67 &        & 4.3      & 4.0     & 2     &                &          \\
21    & 1    & 920,693,907,725       & 16/11/2011 & 14:08:19 & 7.15$\pm$2.01  & 499    & 1.8      & 3.8     & 2     &                &          \\
22    & 1    & 995,340,1004,332      & 16/11/2011 & 14:56:05 & 2.7$\pm$0.11   & 288    & 1.6      & 1.8     & 3     &                &          \\
23    & 1    & 827,662,813,699       & 17/11/2011 & 22:28:37 & 15.36$\pm$0.4  & 365    & 6.4      & 4.3     & 3     & 19.19$\pm$1.55 & E        \\
23    & 2    & 920,744,856,729       & 17/11/2011 & 22:32:49 & 28.19$\pm$0.51 &        & 12.1     & 4.9     & 2.5   &                &          \\
24    & 1    & -881,-588,-910,-549   & 18/11/2011 & 07:34:59 & 17.86$\pm$0.3  & 432    & 14.5     & 15.6    & 3     & 27.43$\pm$4.26 & E        \\
24    & 2    & -848,-608,-901,-572   & 18/11/2011 & 07:29:42 & 16.45$\pm$0.28 & 427    & 21.8     & 23.6    & 3     &                &          \\
24    & 3    & -814,-673,-894,-645   & 18/11/2011 & 07:36:02 & 20.46$\pm$0.58 & 538    & 31.8     & 26.6    & 2     & 35.01$\pm$6.44 & E        \\
25    & 1    & 316,-221,321,-195     & 22/12/2011 & 01:59:34 & 5.13$\pm$0.11  & 156    & 2.0      & 3.0     & 3     & 8$\pm$5        & E,NE     \\
25    & 2    & 272,-141,332,-92      & 22/12/2011 & 01:59:39 & 7.3$\pm$0.16   & 264    & 1.8      & 2.5     & 2.5   &                &          \\
26    & 1    & 1098,13,1126,51       & 16/01/2012 & 00:08:28 & 11.95$\pm$0.13 & 473    & 2.5      & 9.2     & 4.5   & 18.71$\pm$4.5  & E,NE     \\
26    & 2    & 1028,-68,1025,-33     & 16/01/2012 & 00:11:27 & 12.51$\pm$0.19 & 185    & 2.3      & 6.6     & 4     &                & NE       \\
27    & 1    & 1042,93,1072,146      & 09/04/2012 & 01:19:52 & 15.28$\pm$0.4  & 244    & 4.3      & 3.2     & 3     &                &          \\
29    & 1    & -633,339,-628,380     & 08/05/2012 & 13:05:46 & 3.71$\pm$0.05  & 154    & 7.4      & 5.3     & 6     & 7.83$\pm$0.62  & E        \\
31    & 1    & 964,289,945,325       & 26/05/2012 & 20:36:47 & 7.67$\pm$0.04  & 162    & 19.6     & 9.4     & 6     & 24.22$\pm$2.02 & E        \\
31    & 2    & 944,259,944,284       & 26/05/2012 & 20:36:27 & 9.59$\pm$0.09  & 138    & 13.0     & 9.1     & 5     & 17.57$\pm$2.35 & E,NE     \\
31    & 3    & 1116,286,1112,330     & 26/05/2012 & 20:39:53 & 11.56$\pm$0.12 & 532    & 4.5      & 2.7     & 2.5   &                &          \\
32    & 1    & -973,-366,-988,-342   & 30/05/2012 & 08:58:57 & 4.28$\pm$0.02  & 234    & 2.2      & 8.8     & 8     & 15.55$\pm$1.22 & E        \\
32    & 2    & -972,-388,-989,-370   & 30/05/2012 & 08:56:52 & 3.38$\pm$0.02  & 233    & 4.0      & 5.3     & 5     & 19.11$\pm$4.85 & E        \\
33    & 1    & 807,-608,840,-591     & 06/07/2012 & 23:06:45 & 4.69$\pm$0.08  & 314    & 7.0      & 8.6     & 2.5   &                &          \\
33    & 2    & 867,-101,874,-45      & 06/07/2012 & 23:05:07 & 6.52$\pm$0.1   & 407    & 8.5      & 7.7     & 3     &                & E        \\
34    & 1    & -1053,-142,-1076,-129 & 07/08/2012 & 00:59:34 & 9.95$\pm$0.27  & 333    & 15.4     & 7.4     & 2.5   & 16.7$\pm$1.03  & E        \\
35    & 1    & -1024,-148,-1037,-110 & 07/08/2012 & 01:34:47 & 8.78$\pm$0.13  & 327    & 1.9      & 3.0     & 3     &                &          \\
35    & 2    & -1033,-185,-1070,-185 & 07/08/2012 & 01:33:52 & 5.77$\pm$0.1   & 312    & 1.4      & 2.8     & 2.5   &                &          \\
36    & 1    & -1018,-152,-1039,-112 & 07/08/2012 & 03:03:46 & 6.68$\pm$0.1   & 282    & 1.3      & 3.8     & 4     &                &          \\
37    & 1    & -1110,66,-1103,87     & 15/10/2012 & 21:52:17 & 8.27$\pm$0.22  & 358    & 2.3      & 3.8     & 2.5   &                &          \\
38    & 1    & -1046,-180,-1046,-159 & 19/10/2012 & 19:01:00 & 3.04$\pm$0.03  & 224    & 2.7      & 1.6     & 4     &                &          \\
38    & 2    & -1061,-201,-1065,-174 & 19/10/2012 & 19:01:44 & 5.2$\pm$0.08   & 270    & 3.0      & 2.1     & 3     & 15.23$\pm$5.5  & E,NE     \\
38    & 3    & -1121,-189,-1142,-175 & 19/10/2012 & 19:03:00 & 13.08$\pm$0.21 & 424    & 5.2      & 4.9     & 2     &                &          \\
38    & 4    & -1139,-194,-1162,-185 & 19/10/2012 & 19:01:08 & 10.74$\pm$0.18 & 478    & 5.0      & 5.8     & 2     &                &          \\
39    & 1    & -1093,-124,-1140,-107 & 19/10/2012 & 21:04:19 & 10.79$\pm$0.1  & 402    & 6.5      & 4.1     & 4.5   &                & NE       \\
39    & 2    & -1088,-185,-1095,-157 & 19/10/2012 & 21:01:34 & 10.68$\pm$0.12 & 334    & 2.0      & 1.9     & 5     &                & NE       \\
39    & 3    & -1095,-204,-1120,-172 & 19/10/2012 & 21:03:44 & 12.57$\pm$0.36 & 376    & 2.3      & 3.5     & 3.5   &                &          \\
39    & 4    & -1127,-168,-1161,-179 & 19/10/2012 & 21:00:40 & 14.3$\pm$0.17  & 454    & 3.7      & 3.5     & 4     &                & NE       \\
40    & 1    & -1025,-63,-1037,-47   & 20/10/2012 & 18:07:36 & 5.68$\pm$0.06  & 171    & 1.4      & 1.7     & 4     &                &          \\
40    & 2    & -1077,-121,-1065,-96  & 20/10/2012 & 18:09:33 & 5.61$\pm$0.03  & 347    & 9.6      & 4.4     & 7     & 24.83$\pm$3.41 & E,NE     \\
40    & 3    & -1073,-136,-1058,-120 & 20/10/2012 & 18:11:36 & 5.92$\pm$0.7   & 325    & 2.6      & 3.0     & 4     &                &          \\
40    & 4    & -1045,-114,-1020,-110 & 20/10/2012 & 18:10:08 & 5.53$\pm$0.04  & 258    & 3.6      & 2.5     & 6     & 7.32$\pm$1.08  & E,NE     \\
40    & 5    & -1056,-122,-1043,-115 & 20/10/2012 & 18:12:55 & 5.42$\pm$0.02  & 297    & 3.9      & 4.7     & 6     &                &          \\
40    & 6    & -1095,-123,-1079,-101 & 20/10/2012 & 18:09:59 & 6.93$\pm$0.04  & 425    & 6.2      & 3.5     & 11    &                & NE       \\
40    & 7    & -1107,-153,-1094,-121 & 20/10/2012 & 18:11:11 & 5.72$\pm$0.06  & 353    & 3.1      & 3.4     & 12    & 14.17$\pm$2.73 & E,NE     \\
40    & 8    & -1036,-217,-1066,-194 & 20/10/2012 & 18:08:39 & 4.33$\pm$0.08  & 238    & 10.3     & 12.1    & 4.5   & 9.01$\pm$2.16  & E,NE     \\
40    & 9    & -1109,-438,-1117,-399 & 20/10/2012 & 18:11:45 & 6.18$\pm$0.05  & 473    & 12.5     & 13.7    & 3.5   & 13.15$\pm$2.66 & E        \\
40    & 10   & -962,-430,-982,-453   & 20/10/2012 & 18:12:21 & 6.27$\pm$0.03  & 238    & 2.2      & 2.1     & 10    &                & NE       \\
40    & 11   & -978,-404,-999,-418   & 20/10/2012 & 18:11:29 & 4.76$\pm$0.04  & 220    & 1.6      & 1.3     & 4     &                &          \\
43    & 1    & 933,615,894,615       & 07/01/2013 & 06:37:38 & 7.14$\pm$0.07  & 363    & 8.0      & 16.3    & 5     & 7.53$\pm$1.45  & E        \\
43    & 2    & 874,598,890,613       & 07/01/2013 & 06:37:50 & 3.6$\pm$0.03   & 241    & 9.7      & 3.1     & 5     & 9.44$\pm$0.92  & E        \\
43    & 3    & 828,659,816,708       & 07/01/2013 & 06:39:19 & 8.35$\pm$0.08  & 368    & 2.1      & 12.7    & 3.5   & 15.04$\pm$1.81 & E        \\
43    & 4    & 829,606,826,620       & 07/01/2013 & 06:37:01 & 5.16$\pm$0.03  & 222    & 3.9      & 3.5     & 4     &                &          \\
43    & 5    & 801,608,812,631       & 07/01/2013 & 06:37:11 & 4.5$\pm$0.02   & 260    & 1.3      & 2.2     & 5.5   & 14$\pm$2       & E,NE     \\
44    & 1    & 829,644,820,687       & 07/01/2013 & 08:48:37 & 7.23$\pm$0.06  & 295    & 2.8      & 12.3    & 5     & 15.75$\pm$3.09 & E        \\
44    & 2    & 979,637,940,676       & 07/01/2013 & 08:48:17 & 9.78$\pm$0.19  & 512    & 15.8     & 13.4    & 3     & 14.62$\pm$4.96 & E,NE     \\
44    & 3    & 886,644,936,622       & 07/01/2013 & 08:47:19 & 6.95$\pm$0.14  & 352    & 17.2     & 4.8     & 4     & 9$\pm$3        &          \\
44    & 4    & 869,575,879,587       & 07/01/2013 & 08:48:53 & 2.41$\pm$0.05  & 202    & 2.9      & 2.8     & 4     &                &          \\
45    & 1    & -396,367,-409,379     & 17/02/2013 & 15:45:42 & 2.48$\pm$0.04  & 92     & 0.7      & 3.1     & 3.5   & 7.82$\pm$1.66  & E        \\
46    & 1    & -1024,-281,-1038,-268 & 24/05/2013 & 18:55:12 & 12.07$\pm$0.23 & 430    & 1.5      & 4.4     & 2.5   &                &          \\
46    & 2    & -1102,-389,-1080,-363 & 24/05/2013 & 18:53:58 & 10.99$\pm$0.11 & 498    & 2.4      & 3.7     & 5     &                & NE       \\
46    & 3    & -1032,-332,-1054,-327 & 24/05/2013 & 18:54:34 & 9.9$\pm$0.1    & 384    & 2.5      & 3.8     & 6     &                & NE       \\
47    & 1    & 207,-251,223,-243     & 27/05/2013 & 01:53:28 & 5.27$\pm$0.14  & 225    & 0.6      & 1.6     & 3     &                &          \\
47    & 2    & 237,-197,270,-188     & 27/05/2013 & 02:02:59 & 5.02$\pm$0.12  & 222    & 2.9      & 1.6     & 3     &                &          \\
48    & 1    & -1076,77,-1044,111    & 18/07/2013 & 17:59:56 & 15.28$\pm$0.16 & 540    & 12.3     & 22.0    & 3.5   & 21.98$\pm$15.6 & E,NE     \\
48    & 2    & -1134,36,-1069,99     & 18/07/2013 & 17:59:01 & 15.76$\pm$0.12 & 588    & 25.4     & 27.4    & 5     & 26.64$\pm$2.17 & E,NE     \\
48    & 3    & -1153,41,-1102,112    & 18/07/2013 & 17:58:34 & 16.08$\pm$0.21 & 597    & 19.9     & 23.7    & 4     & 15.76$\pm$3.09 & E        \\
48    & 4    & -1084,-59,-1069,-33   & 18/07/2013 & 17:56:10 & 9.23$\pm$0.23  & 426    & 7.7      & 7.6     & 4     &                &          \\
48    & 5    & -1139,-15,-1102,42    & 18/07/2013 & 17:57:31 & 15.83$\pm$0.21 & 471    & 17.9     & 13.0    & 3.5   &                &          \\
49    & 1    & -1041,93,-1117,118    & 11/10/2013 & 07:12:01 & 15.12$\pm$0.47 & 484    & 13.6     & 22.9    & 2     &                &          \\
49    & 2    & -1030,-126,-1058,-106 & 11/10/2013 & 07:15:17 & 7.73$\pm$0.14  & 197    & 3.6      & 6.4     & 3     &                &          \\
49    & 4    & -1044,409,-1082,429   & 11/10/2013 & 07:13:30 & 10.45$\pm$0.17 & 386    & 17.9     & 8.0     & 3     & 15.38$\pm$2.58 & E        \\
49    & 5    & -1020,-76,-1071,-51   & 11/10/2013 & 07:15:51 & 8.03$\pm$0.18  & 191    & 10.4     & 13.0    & 3     & 9.37$\pm$1.22  & E        \\
52    & 1    & -710,65,-721,113      & 04/01/2014 & 15:32:47 & 5.93$\pm$0.12  & 183    & 3.0      & 12.5    & 3     &                &          \\
53    & 1    & 1101,-296,1136,-291   & 06/01/2014 & 07:42:35 & 9.48$\pm$0.22  & 420    & 1.5      & 8.4     & 1.5   &                &          \\
54    & 1    & 1115,220,1143,254     & 10/02/2014 & 21:00:57 & 8.33$\pm$0.07  & 408    & 1.3      & 3.2     & 3     &                &          \\
54    & 2    & 1118,55,1134,82       & 10/02/2014 & 21:01:06 & 7.46$\pm$0.1   & 400    & 4.2      & 8.0     & 3     &                &          \\
54    & 3    & 1062,110,1078,106     & 10/02/2014 & 21:02:02 & 2.32$\pm$0.05  & 238    & 1.0      & 0.9     & 3     &                &          \\
54    & 4    & 1108,29,1138,43       & 10/02/2014 & 20:58:27 & 3.77$\pm$0.13  & 355    & 7.5      & 3.0     & 2     &                &          \\
54    & 5    & 1076,-6,1091,1        & 10/02/2014 & 20:59:10 & 4.8$\pm$0.1    & 257    & 3.6      & 2.9     & 4.5   & 19.72$\pm$3.23 & E,NE     \\
55    & 1    & 1123,231,1143,244     & 10/02/2014 & 22:48:05 & 8.63$\pm$0.24  & 405    & 5.9      & 3.7     & 2.5   &                &          \\
55    & 2    & 1167,190,1203,192     & 10/02/2014 & 22:44:59 & 6.54$\pm$0.17  & 477    & 1.9      & 2.8     & 3     &                &          \\
56    & 1    & 1124,-8,1186,1        & 11/02/2014 & 13:28:08 & 9.07$\pm$0.14  & 403    & 9.3      & 7.8     & 4     & 20.71$\pm$4.71 & E,NE     \\
56    & 2    & 1068,-39,1130,-6      & 11/02/2014 & 13:28:13 & 11.88$\pm$0.13 & 314    & 17.8     & 27.6    & 5     & 19.62$\pm$2.96 & E        \\
56    & 3    & 1027,82,1043,54       & 11/02/2014 & 13:26:19 & 3.22$\pm$0.16  & 205    & 16.5     & 9.2     & 2.5   &                &          \\
56    & 4    & 1062,397,1122,348     & 11/02/2014 & 13:27:33 & 14.38$\pm$0.34 & 501    & 16.2     & 23.2    & 2     &                &          \\
56    & 5    & 1026,332,1004,354     & 11/02/2014 & 13:28:38 & 13.5$\pm$0.16  & 431    & 6.9      & 10.7    & 4.5   & 24.17$\pm$5.13 & E        \\
56    & 6    & 998,318,979,345       & 11/02/2014 & 13:28:49 & 7.59$\pm$0.2   & 392    & 3.6      & 3.8     & 2.5   &                &          \\
56    & 7    & 1068,427,1023,456     & 11/02/2014 & 13:31:32 & 14.16$\pm$0.55 & 457    & 4.9      & 15.0    & 3     & 13.64$\pm$3.93 & E,NE     \\
56    & 8    & 1015,384,990,419      & 11/02/2014 & 13:30:19 & 10.64$\pm$0.15 & 379    & 3.5      & 9.0     & 4     &                &                
\label{table1}
\end{longtable}
\end{landscape}

\twocolumn
\end{document}